\documentclass[%
 reprint,
 amsmath,amssymb,
 aps,
]{revtex4-2}

\usepackage{graphicx}
\usepackage{dcolumn}
\usepackage{bm}

\begin{document}

\preprint{APS/123-QED}

\title{Bistable Fourth Sound Resonance in Superfluid $^3$He-B due to Gap Suppression}

\author{Alexander J. Shook}
\email{ashook@ualberta.ca}
\author{Daksh Malhotra}%
\author{Aymar Muhikira}%
\author{John P. Davis}%
\email{jdavis@ualberta.ca}
\affiliation{%
 Department of Physics, University of Alberta, Edmonton, Alberta T6G 2E1, Canada}%

\date{\today}

\begin{abstract}

Superfluidity in $^3$He exhibits many unique properties that are of interest to modern condensed matter research, including multiple superfluid phase transitions, topological defects, and exotic classes of excitations like Majorana and Weyl fermions. Many of the most interesting theoretical proposals, which remain underexplored, are realized in highly confined geometries, where surface effects play a dominant role in the thermodynamic and hydrodynamic properties. We have developed nanofluidic resonators capable of exciting a fourth-sound acoustic mode in thin channels with a highly confined dimension ($750-1800$ nm) that is only $1-2$ orders of magnitude larger than the superfluid coherence length. When a sufficiently large drive force is applied, we observe a non-linear softening of the resonance that we interpret as due to the flow suppression of the superfluid gap. We have developed a model of the device that allows the resonance amplitude to be calibrated into a superfluid velocity, which exhibits critical behavior at particular velocities. We identify one of the observed critical velocities as being the velocity at which the gap component parallel to the flow is suppressed to zero. We compare the calibrated velocity to the prediction of a Ginzburg-Landau model, and find reasonable agreement. This measurement represents an ongoing effort to link the hydrodynamic measurements of these nanofluidic devices to theoretical predictions regarding surface gap suppression and surface-bound states.

\end{abstract}

\maketitle

\section{Introduction}

Superfluid $^3$He is an exotic p-wave, spin-triplet superfluid exhibiting a rich symmetry group that allows for many superfluid phases. Only two phases ($^3$He-A and $^3$He-B) are thermodynamically stable in bulk systems without external fields \cite{vollhardt2013superfluid}. The existence of boundaries, however, leads to surface pair-breaking, which suppresses the component of the superfluid gap that is transverse to the surface \cite{li1988superfluid,thuneberg1998models,wiman2014superfluid,rudd2021strong,saraj2025dimensional}. The effect of this gap suppression on the phase diagram of superfluid $^3$He can be demonstrated within the theoretical framework of Ginzburg-Landau theory and has a pronounced effect when one or more dimensions of the system become comparable to the coherence length \cite{li1988superfluid}. In addition to modifying the stability of various phases, the presence of walls leads to the existence of localized bound states. These are low-energy excitations that cannot exist in the bulk superfluid due to its gapped nature, but can exist close to a surface (or other defects) \cite{nagai2008surface}. Such bound states are interesting in their own right due to the expectation that some $^3$He bound states may be Majorana \cite{okuda2012surface,tsutsumi2012edge,wu2013majorana,wu2013majorana,park2015surface} or Weyl fermions \cite{stone2016effective,shevtsov2016electron,volovik2017chiral}. Confined systems may also help stabilize topological defects such as quantized vortices or domain walls, which may not exist in the bulk fluid \cite{salomaa1987quantized,yamashita2008spin,kawakami2009stability,shook2024topologically}.

To study these unique properties, experimental advances have been made in both confining and probing nanoscale superfluid $^3$He \cite{levitin2013phase,zhelev2017ab,levitin2019evidence,shook2020stabilized,shook2024surface,shook2024topologically}. One of the traditional methods for investigating the properties of superfluids is to study their flow properties. Such experiments date back to the discovery of superfluidity \cite{kapitza1938viscosity}, which led to the discovery of novel hydrodynamic phenomena such as the thermo-mechanical fountain effect \cite{allen1938new,london1939thermodynamics}. In this instance, the experiment \cite{allen1938new} measured a qualitatively new macroscopic effect which motivated the development of two-fluid hydrodynamics \cite{tisza1938transport,landau2018theory}. In other cases, however, the purpose of flow experiments in superfluids is to measure hydrodynamic or thermodynamic variables that yield insights into the microscopic origins of superfluidity. 

A classic example that links macroscopic flow to microscopic excitations is the argument outlined by Landau for the origins of inviscid flow in superfluids \cite{landau1947theory}. The crux of the argument was to notice that for a frictional force to exist between a fluid and the surface it is flowing over (necessary for the creation of viscosity), thermal excitations must be created at the surface, which dissipate the total momentum of the fluid into heat. Unlike classical fluids, superfluid excitations are gapped, implying a minimum energy scale and, therefore, a minimum velocity required to create these excitations. There exists a critical velocity, defined by the lowest energy gapped excitations, which separates an inviscid superfluid regime from a viscous regime. 

In superfluid $^3$He this energy gap corresponds to the energy required to break the Cooper pairs that comprise the superfluid condensate. Unlike s-wave superconducting systems, this energy gap may be anisotropic in momentum space, and the information about this anisotropy is typically encoded in a $3\times 3$ matrix order parameter \cite{vollhardt2013superfluid}, with temperature and pressure dependence that may be described within a Ginzburg-Landau framework \cite{li1988superfluid,yapa2022triangular,sun2023superfluid,saraj2025dimensional}. Measurements of critical velocities in superfluid $^3$He have revealed many subtle aspects of this picture not captured by the simple argument of Landau. Experiments using vibrating wires and other mechanical oscillators have shown that the true critical velocity is a fraction (determined in part by the geometry of the oscillator) of the Landau critical velocity \cite{castelijns1986landau,bradley2016breaking}. It has been argued that the reason for this discrepancy is the suppression of the superfluid gap at the surface of the oscillator, or in other words, the existence of surface-bound states -- which can have Majorana- or Weyl-like nature. When the superfluid flows over a surface with bound states, the dispersion relation of all excitations (bound or otherwise) is shifted by the Galilean transformation. When the flow velocity is high enough, it may be energetically possible for these bound states to escape into the bulk fluid. For alternating flow over a surface, a bound state pumping process occurs, which produces a suppressed critical velocity \cite{bradley2016breaking,shook2024surface}.

These mechanical oscillator experiments demonstrate a direct link between macroscopic hydrodynamic measurements and subtle aspects of the underlying microscopic system. Theoretical interest in the role of surfaces in modifying the superfluid and its elementary excitations motivates the creation of highly confined systems where surface effects become dominant. For example, an intriguing theoretical proposal by Wu and Sauls \cite{wu2013majorana}, is to study DC flow of superfluid $^3$He-B confined between parallel plates separated by tens of coherence lengths. At sufficiently low temperatures, the contribution to the total mass current due to Majorana bound states is expected to display a characteristic $T^3$ temperature scaling. Such a measurement would demonstrate the existence of Majorana fermions, which have long been predicted to exist in condensed matter systems, but have remained experimentally elusive \cite{mourik2012signatures,williams2012unconventional,rokhinson2012fractional,deng2012anomalous,das2012zero,finck2013anomalous,churchill2013superconductor,lee2014spin,deng2016majorana,nichele2017scaling,chen2017experimental,gul2018ballistic,grivnin2019concomitant,choy2011majorana,pientka2013topological,nadj2014observation,pawlak2016probing,feldman2017high,kim2018toward,pawlak2019majorana,menard2017two,palacio2019atomic,kayyalha2020absence}.

\begin{figure}[t!]
    \centering
    \includegraphics[width=\linewidth]{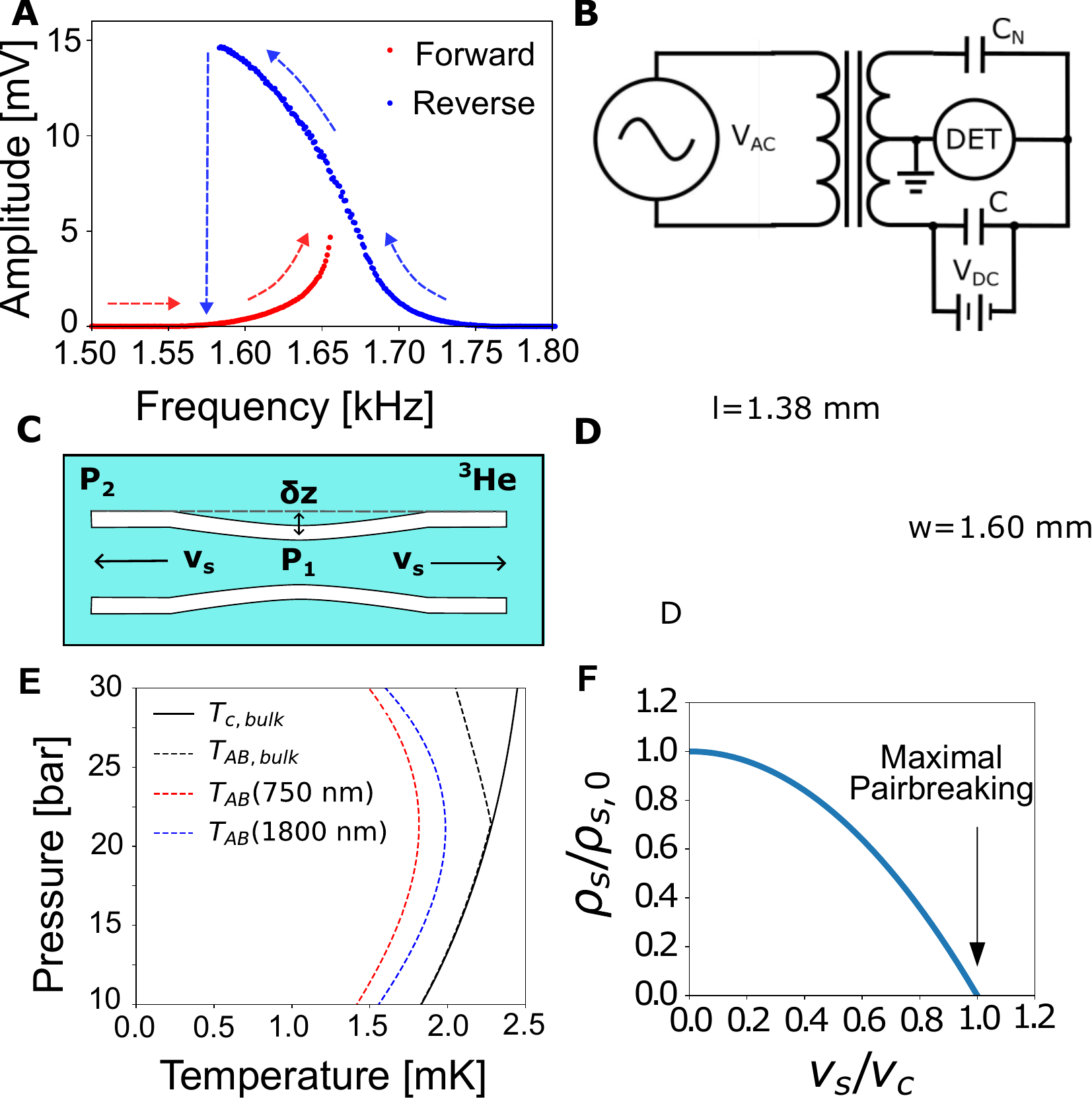}
    \caption{a) Example of a bistable frequency response curve measured by sweeping the drive frequency of the Helmholtz resonator. Due to the hysteretic nature of the bistability, the measured branch depends on the initial frequency when sweeping through the resonance. Here, red data points correspond to the lower branch measured by sweeping the frequency from low to high (dubbed the ``forward" frequency sweep), and blue data points show a frequency sweep from high to low frequencies (``reverse" sweep). b) Simplified diagram of the bridge circuit used to measure the time-varying capacitance signal from the Helmholtz resonator. c) Simplified cross-section of a Helmholtz resonator device where the basin has been compressed by $\delta z$. This has the effect of increasing the fluid pressure to $P_1$, which differs from the external pressure by $\delta P = P_2-P_1$, which drives motion in the fluid with velocity $v_s$. d) Diagram of the confined volume of the Helmholtz resonator with dimensions. e) Phase diagram of superfluid $^3$He. The solid black line denotes the superfluid critical temperature, and the dashed black line the bulk A to B transition curve. The red and blue dashed lines denote the A-B transition line for a parallel plate geometry with confined dimensions of 750 and 1800 nm respectively. f) Plot of the reduction in superfluid density due to velocity-dependent gap suppression. Here, a simple parabolic dependence is assumed, which does not account for the multiple order parameter components of the planar distorted B-phase. The velocity at which $v_s=v_c$, such that $\rho_s = 0$, is the maximal pair-breaking critical velocity.}
    \label{fig:bistability}
\end{figure}

To study the flow properties of $^3$He confined to a quasi-2D parallel plate geometry, we have developed nanofluidic resonators, known as Helmholtz resonators, which have been described elsewhere \cite{shook2020stabilized,shook2024surface}. A confined volume with well-defined geometry is created by etching quartz chips using optical lithography, which are then bonded together. A drawing of the confined volume can be seen in Figure~\ref{fig:bistability}. The central circular region is referred to as the basin, which is connected to an external reservoir of $^3$He via two channels. The height of the basin and channels, $D$, ranges from hundreds of nanometers to a few microns for typical devices. Aluminum electrodes are patterned onto the surfaces to create a parallel plate capacitor that is used to both drive the system and perform a capacitive readout. When the capacitor is driven with an alternating voltage close to an acoustic resonance, a fourth sound mode is excited inside the channels of the device. In past experiments, we have demonstrated the ability to map out phase diagrams for the confined system by studying the temperature dependence of the resonant frequency \cite{shook2020stabilized}. We have also performed characterizations of the velocity-force curves for these devices, which led to the observation of a critical velocity in the A-phase due to bound state excitations, which is not observed in bulk $^3$He-A due to textural effects \cite{shook2024surface}. Here, we report on critical velocity measurements in the B-phase, which exhibit qualitatively different behavior from the A-phase. 

We studied two devices with confined dimensions of 1800 and 750 nm. The Helmholtz resonators are cooled in an experimental cell of pressurized $^3$He to a temperature in the B-phase region of the pressure-temperature phase diagram (see Figure \ref{fig:bistability}e). Once prepared in the desired state, frequency sweep measurements are carried out at various drives, exciting the Helmholtz resonance described above. Using a hydrodynamic model (see Section II), we can calibrate the driving voltage into a pressure gradient down the channels, and the amplitude of the measured resonance into a superfluid mass flux. Unlike the measurement in the A-phase \cite{shook2024surface}, where the peak shape flattens at high drives, in the B-phase, we observe a qualitatively different non-linear response. Namely, we observe a bistable, hysteretic, resonance reminiscent of a Duffing oscillator frequency response. An example of a characteristic frequency response curve can be seen in Figure \ref{fig:bistability}.

The plot shows the two measurable branches of the bistable resonance, which are measured by sweeping the frequency either from low to high frequencies (the ``forward" sweep), or in the opposite direction (the ``reverse" sweep). The reverse sweep direction leads to a ``shark-fin" like shape that is characteristic of a resonator with a non-linear restoring force. In our case, where the curve shifts to the left of the low-drive resonant frequency, this would be an instance of non-linear softening. We emphasize that such a bistable resonance is not observed in otherwise identical measurements in $^3$He-A \cite{shook2024surface}, or similar measurements in $^4$He in the same drive regime \cite{varga2020observation}. This confirms that a material property unique to the B-phase is producing this non-linear softening effect. We suggest that this softening is best explained by the flow-dependent suppression of the superfluid gap, which is captured most simply by a Ginzburg-Landau model. This model posits a non-linear relationship between the superfluid velocity $v_s$ and superfluid mass flux $j_s$. We show that introducing this non-linear dependence into the hydrodynamic equations describing our devices leads to a non-linear equation of motion with solutions that resemble a Duffing oscillator for drives that are not too large.

By studying how the amplitude of these resonances scales with drive strength, we are able to identify discontinuities in the slope that correspond to two distinct critical velocities. At low velocities, there exists a threshold where the velocity-force scaling first becomes non-linear. The second critical velocity occurs at velocity scales an order of magnitude larger. Such a critical velocity is expected to exist within the framework of the Ginzburg-Landau theory when the flow is large enough to cause the order parameter component parallel to the flow to be suppressed to zero. We compare the critical velocity calculated from our measurements to the prediction of the GL theory, and find them to be in reasonable agreement. We contrast this measurement to the A-phase, where a critical velocity due to surface-bound state pumping is observed \cite{shook2024surface}, but no bistability is present.


\section{Model}

To explain the bistability observed in our measurements, it is first necessary to describe the hydrodynamic model of a Helmholtz resonator. When the capacitor plates are driven, an electrostatic force is created between the plates, which compresses the basin. This force is opposed by the stiffness of the quartz substrate and any pressure difference between the basin and the fluid outside the Helmholtz resonator. Given an effective mass of $m_p$, the equation of motion for the plate deformation is
\begin{equation}
    m_p \delta \ddot{\bar{z}} = F_{E} - k_p \delta \bar{z} - A \delta \bar{P},
\end{equation}
where $k_p$ is the effective stiffness of the quartz, $F_E$ is the electrostatic force between the electrodes, $A$ is the area of the basin, $\delta \bar{P}$ is the difference between the spatially averaged pressure inside the basin and the pressure outside the basin, and $\delta \bar{z}$ is the mean plate deflection. For a clamped circular plate, the mean deflection is expected to be one-third of the maximum plate deflection. When the system is driven with no fluid in the device (such that $\delta \bar{P} = 0$), the plate deflection oscillates resonantly at the frequency $\omega_p = \sqrt{k_p/m_p}$, which is on the order of 100 kHz for typical devices in vacuum, and greater than 10 MHz when submersed in liquid $^3$He. When the Helmholtz resonator is submerged in fluid, the plate mode becomes coupled to a hydrodynamic mode via the pressure term in the equation of motion. The resonant frequency for this hydrodynamic mode is typically a few kHz or smaller. When the drive frequency of the coupled system is small compared to $\omega_p$ (always appropriate to our mode of operation), the inertial term in Equation 1 can be neglected, such that
\begin{equation}
    \delta \bar{z} = \frac{F_E-A\delta P}{k_p}.
    \label{eq:plate_deflection}
\end{equation}
When the basin wall is deflected from its equilibrium position, the change in volume is 
\begin{equation}
    V_B(t) = AD \left(1-2\frac{\delta \bar{z}(t)}{D}\right),
\end{equation}
where $D$ is the separation between the plates for zero electrostatic force. The time derivative of the volume is related to the change in basin mass via the chain rule
\begin{equation}
    \dot{M}_B = \rho_0 A D \left(\frac{\delta \dot{\bar{\rho}}}{\rho_0} - 2 \frac{\delta \bar{z}}{D} \right).
    \label{eq:mass_chain_rule}
\end{equation}
Here, we assume that the fluctuations in density $\delta \bar{\rho}$ are small compared to the mean density $\rho_0$. This change in mass couples to fluid flow via the mass continuity equation, which states that the time derivative of the total mass in the basin, $M_B$, must be equal to the mass flux out of the channels
\begin{equation}
    \dot{M}_B = - 2 \oint \vec{j} \cdot d\vec{a} = -2 \langle j_s \rangle a,
\end{equation}
where $a$ is the cross-sectional area of a channel. The vector $\vec{j}$ is the mass flux through a single channel, and the factor of $2$ accounts for the fact that there are two identical channels. Here, the angled brackets denote a spatial average across the channel cross-section. We consider a simple model where the mass flux vector may be taken to be parallel to the cross-section normal everywhere, neglecting the converging flow at the mouth of the channel \cite{shook2024surface}. This converts the system into a one-dimensional hydrodynamic problem. 

The pressure and density fluctuations can be equated using the isothermal compressibility $\kappa_T$, or equivalently the speed of first sound $c_1$,
\begin{equation}
    \delta P = \frac{1}{\kappa_T} \frac{\delta \rho}{\rho_0} = c_1^2 \delta \rho.
    \label{eq:linear_acoustic_eq}
\end{equation}
Any pressure gradient down the channels is expected to drive superfluid flow according to Euler's equation (neglecting the advection term)
\begin{equation}
    |\dot{\vec{v}}_s| = \frac{|\vec{\nabla}P|}{\rho_0} \approx \frac{\delta \bar{P}}{\rho_0 \ell}.
\end{equation}
Here $\ell$ is the length of the channel, over which the basin pressure changes to the pressure outside the device. Putting together equations \ref{eq:plate_deflection} and \ref{eq:mass_chain_rule}-\ref{eq:linear_acoustic_eq} gives the equation 
\begin{equation}
    \ddot{v}_s + \left(\frac{\Sigma}{1+\Sigma} \right) \left(\frac{2ac_1^2}{\rho A D \ell}\right) \langle j_s \rangle = \left(\frac{1}{1+\Sigma} \right) \frac{\dot{F}_E}{\rho_0 A \ell},
    \label{eq:intermediate_eq}
\end{equation}
where we have defined the ratio $\Sigma = k_pD/2\rho_0 A c_1^2$. This nearly has the form of a harmonic oscillator, but to make the form explicit, we must establish the relationship between $\langle j_s \rangle$ and $v_s$. The proportionality factor is the spatially averaged superfluid density given by
\begin{equation}
    \langle j_s \rangle = \langle \rho_s \rangle v_s.
\end{equation}
For low flow velocities, $\langle \rho_s \rangle$ is expected to be a constant, but in the bound state dissipation velocity regime, it may be a function of velocity such that $j_s(v_s)$ is non-linear. Equation \ref{eq:intermediate_eq} can then be simplified to
\begin{equation}
    \ddot{v}_s + \omega_0^2 v_s = \left(\frac{1}{1+\Sigma} \right) \frac{\dot{F}_E}{\rho_0 A \ell},
\end{equation}
by identifying the resonant frequency
\begin{equation}
    \omega_0^2 = \frac{1}{1+\Sigma} \left(\frac{k_p a}{\rho A^2 \ell} \right) \frac{\langle \rho_s \rangle}{\rho} = \left(\frac{2\Sigma}{1+\Sigma} \right) q_H^2 c_4^2,
    \label{eq:resonance_freq}
\end{equation}
where $c_4 = \sqrt{\langle \rho_s \rangle/\rho_0} c_1$ is the speed of fourth sound, and $q_H = \sqrt{a/AD\ell}$ is the standard Helmholtz wavenumber determined by the geometry. The quantity $q_H c_4$ can be understood as the frequency of a pure Helmholtz resonance uncoupled to the plate motion, and the ratio $2\Sigma/(1+\Sigma)$ represents a shift to the frequency due to the plate coupling. Written in this form, it is clear that the resonance may become non-linear when $\langle \rho_s \rangle$ is velocity dependent. Such velocity dependence is expected due to the flow suppression of the superfluid gap (see Figure \ref{fig:bistability}f). This effect may be described within the framework of Ginzburg-Landau theory for temperatures close to the critical temperature. A description of this theory applied to our system can be found in Appendix A. 

When gap suppression occurs, the first-order correction to $\rho_s$ is parabolic
\begin{equation}
    \langle \rho_s \rangle = \rho_{s0} \left(1 - \frac{v_s^2}{v_c^2} \right).
    \label{eq:rhos_vs_scaling}
\end{equation}
Here $v_c$ is the ``total pair-breaking" critical velocity, where all components of the order parameter are suppressed to zero, and $\rho_s \to 0$. The pre-factor $\rho_{s0}$ is the superfluid density for the static fluid. It should be noted that by using this velocity-dependent expression for $\langle \rho_s \rangle$, we are assuming that the system is always close to thermal equilibrium during the experiment. 

Substituting this expression into Equation~\ref{eq:intermediate_eq}, yields an expression very similar to the Duffing oscillator. To make the form more clear, we introduce the variables $x = v_s/v_c$, $G = (1+\Sigma)^{-1}(F_E/\rho_0A\ell v_c)$, and $\omega_{00}^2 = 2\Sigma (1+\Sigma)^{-1} q_H^2 c_1^2 (\rho_{s0}/\rho_0)$, and add a phenomenological damping term, $\nu \dot{x}$, such that
\begin{equation}
    \ddot{x} + \nu \dot{x} + \omega_{00}^2(1-x^2)x = \dot{G}.
    \label{eq:Duffing}
\end{equation}
The only difference between this expression and the general form of a driven damped Duffing oscillator \cite{landau2013mechanics}
\begin{equation}
    \ddot{x} + \nu \dot{x} + \omega_{00}^2 x + \kappa x^3 = F/m,
\end{equation}
is that we have the particular case where the cubic term is the negative of the linear regime resonant frequency $\kappa = - \omega_{00}^2$, and the right-hand side features a time derivative in the driving term (note that $G$ has units of frequency). This is a consequence of the fact that the differential equation is written in terms of velocities rather than displacements.

Equation~\ref{eq:Duffing} can be solved using a harmonic balance method \cite{krack2019harmonic}. By assuming a sinusoidal drive force $G = G_0 \sin(\omega t)$, and an ansatz of the form
\begin{equation}
    x(t) = x_0 \cos(\omega t + \theta),
    \label{eq:ansatz}
\end{equation}
we find the corresponding expression for the cubic term is
\begin{equation}
    x^3(t) = \frac{3}{4}x_0^3\cos(\omega t + \theta) + \frac{1}{4} x_0^3 \cos(3\omega t + 3 \theta).
\end{equation}
The simplest version of the harmonic balance method consists in substituting Equation~\ref{eq:ansatz} into the differential equation, then dropping all terms oscillating at frequency multiples greater than $1\omega$. One can then derive two equations specifying the amplitude and phase of the oscillator, as functions of the frequency
\begin{equation}
    \left[\omega_{00}^2\left(1-\frac{3}{4}x_0^2\right) - \omega^2 \right]^2 + (\nu \omega)^2 = \left(\frac{\omega G_0}{x_0} \right)^2,
\end{equation}
\begin{equation}
    \tan(\theta) = \frac{\omega_{00}^2\left(1-\frac{3}{4}x_0^2\right) - \omega^2}{\nu \omega}.
\end{equation}
The equation for $x_0(\omega)$ must necessarily be written in implicit form, because it is not a single-valued function. It is instructive to point out that the maximum value of each curve coincides with a parabolic ``backbone" curve, Figure 2a. This curve can be found by taking $G_0/x_0 \to 0$ and $\nu \to 0$, such that
\begin{equation}
    \frac{\omega_b^2}{\omega_{00}^2} = 1 - \frac{3}{4}x_{b}^2 = 1 - \frac{3}{4}\frac{v_{sb}^2}{v_c^2}.
    \label{eq:backbone}
\end{equation}
This backbone curve provides a useful method for comparing the model to our measurements. By extracting the frequency and amplitude of the discontinuity, one can construct a backbone curve $\omega_b^2(x_b^2)$. Comparison to the quadratic form in Equation~\ref{eq:backbone} allows us to check if the bistable resonance follows the drive scaling expected of a Duffing oscillator and to extract the value of $v_{c}$ from the fit, Figure 2b. 
\begin{figure}
    \centering
    \includegraphics[width=\linewidth]{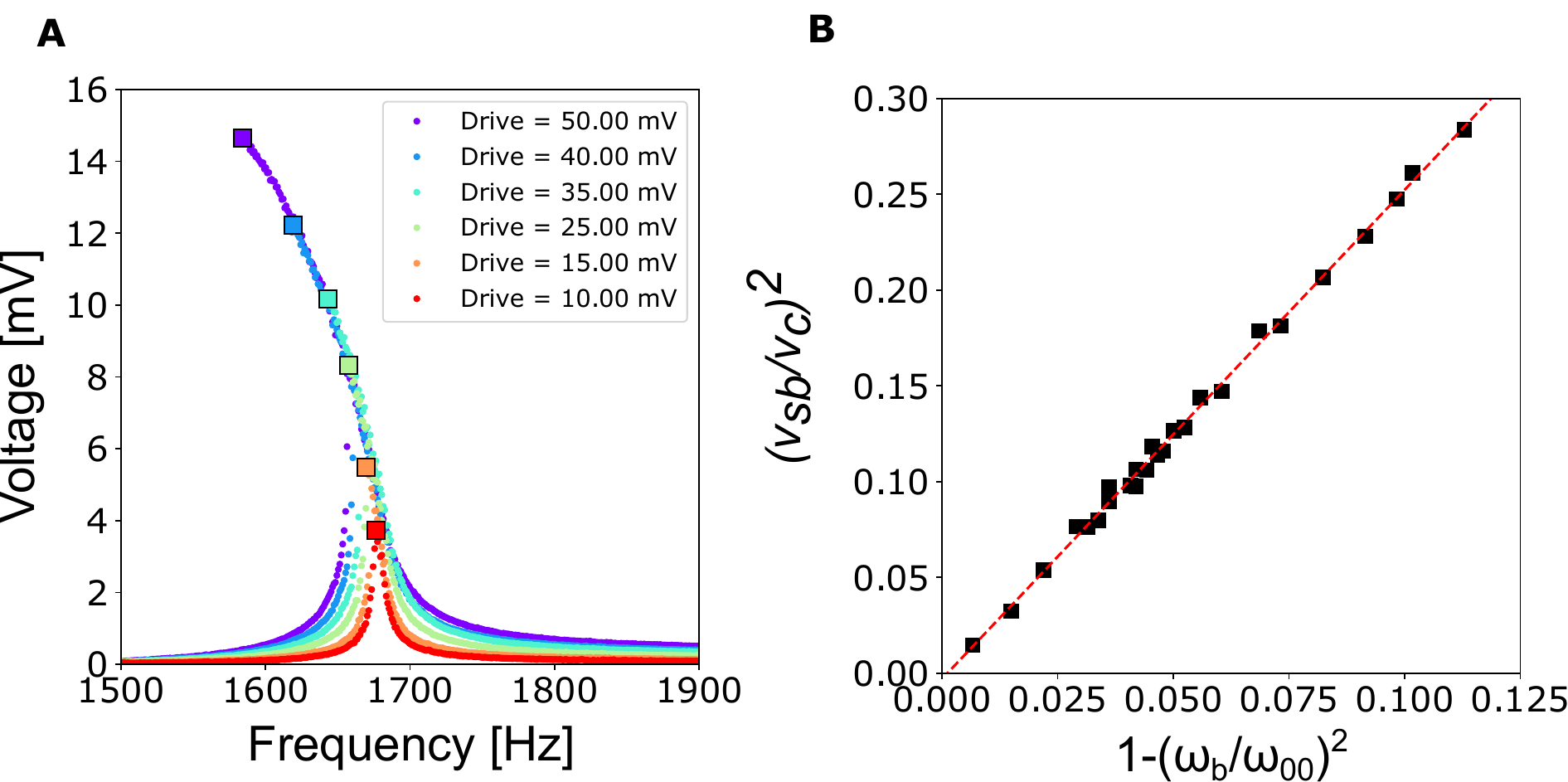}
    \caption{a) Plot of the non-linear drive scaling of the 1800 nm device at 25 bar and $T = 0.72 T_c$ for drive voltages of 10-50 mV. The maximum values are marked with squares in a color corresponding to the drive voltage in the legend. For each drive voltage, two parameters are extracted: the maximum amplitude, and the frequency at which the maximum amplitude occurs $f_c$. These coordinates trace the backbone of the non-linear resonance. b) The maximum amplitude of the voltage resonance is converted into a mass flux using Equation~\ref{eq:vcal}. This is then converted to normalized by the critical velocity expected by the GL theory $v_{sb}/v_c = (2m\xi/\hbar)v_{sb}$. The square of this normalized velocity is then plotted against $1-(\omega_{b}/\omega_{00})^2$, to test the relationship predicted by the Duffing like backbone equation seen in equation \ref{eq:backbone}.}
    \label{fig:drive_scaling}
\end{figure}

The final step in comparing the measurement to this model is to establish the coupling of the superfluid velocity to the capacitive readout. The Helmholtz capacitance is time varying due to the deflections in the capacitor plates. Because $\delta z \ll D$, the expression can be linearized
\begin{equation}
    C(t) = \frac{\epsilon A}{D-2\delta z(t)} \approx C_0 \left( 1 + 2\frac{\delta z(t)}{D} \right), \quad C_0 = \frac{\epsilon A}{D}.
    \label{eq:capacitance}
\end{equation}
The Helmholtz resonator is connected to a capacitance bridge circuit shown in Figure~\ref{fig:bistability}b. When the reference capacitance of the bridge is tuned to $C_0$, the current into the detector fluctuates around zero. The current from the capacitor is given by $\frac{d}{dt} (CV)$, where both the capacitance and the voltage are time varying. This produces signals that oscillate at $1\omega$, $2\omega$, and $3\omega$. We make use of a lock-in amplifier to demodulate the signal at $1\omega$, such that we measure a signal
\begin{equation}
    V_{\textrm{DET}}^{(1\omega)} = R_{\textrm{tran}} \dot{C} V_{\textrm{DC}},
\end{equation}
where $V_{\textrm{DC}}$ is the bias applied to the Helmholtz resonator, and $R_{\textrm{tran}}$ is the effective resistance of the transimpedance amplifier we use to convert the current from the capacitor into a voltage. Using the linearized capacitance expression in Equation~\ref{eq:capacitance}, we define the normalized quantity
\begin{equation}
    u = \frac{1}{\tau} \frac{V_{\textrm{DET}}^{(1\omega)}}{V_{\textrm{DC}}} = \frac{\delta \dot{\bar{z}}}{D}, \quad \tau = 2R_{\textrm{tran}}C_0,
\end{equation}
which is a normalized plate velocity. Using the balance of forces on the plate and Euler's equation, we can write
\begin{equation}
    u = \frac{\dot{F}_E}{k_p D} - \frac{\rho_0 A \ell}{k_p D} \ddot{v}_s.
\end{equation}
The driving force can be eliminated in favor of the superfluid mass flux using Equations~\ref{eq:mass_chain_rule}-\ref{eq:intermediate_eq} (neglecting the damping term) to show that
\begin{equation}
    u = \frac{\rho_s a v_s}{\rho A D} + \Sigma \frac{\rho_0 A \ell}{k_p D} \ddot{v_s}
    \label{eq:u_full}
\end{equation}
Using the harmonic balance approximation $\ddot{v}_s \to -\omega^2 v_s$, this can be written as 
\begin{equation}
    u = \frac{[(1+\Sigma) \omega_0^2 - \Sigma \omega^2)]}{\Sigma} \frac{\ell v_s}{2c_1^2}
\end{equation}
Since the compressibility term is small, $\Sigma \ll 1$, the frequency is near resonance, the second term can be neglected for drives that are not too large. In this case $u$ is proportional to $\langle j_s \rangle = \langle \rho_s \rangle v_s \propto \omega_0^2 v_s$. For larger drives, this assumption breaks down, as discussed in Appendix D. 

Here, we consider only the weakly non-linear regime, where the second term can be neglected, such that
\begin{equation}
    \langle j_s \rangle = \frac{\rho_0AD}{a} \frac{1}{\tau} \frac{V_{\textrm{DET}}^{(1\omega)}}{V_{\textrm{DC}}}.
    \label{eq:current_calibration}
\end{equation}
This means we can convert the amplitude of the detector signal (in voltage) into a measured mass flux. If we instead want a measurement of velocity, we must make use of both the amplitude and the amplitude-dependent frequency $\omega_0^2$
\begin{equation}
    v_s = \frac{\langle j_s \rangle}{\langle \rho_s \rangle} = \frac{\omega_{00}^2}{\omega_0^2} \frac{\langle j_s \rangle}{\rho_{s,0}} = \left(\frac{k_p D}{\rho_0 A \ell} \right) \frac{1}{\omega_0^2 \tau} \frac{V_{\textrm{DET}}^{(1\omega)}}{V_{\textrm{DC}}}.
    \label{eq:vcal}
\end{equation}
To check if this calibrated velocity obeys drive scaling expected by the Duffing-like model, we can perform an experiment where we sweep the frequency of the oscillator for various fixed drives. For each reverse sweep, we can extract a maximum amplitude and a frequency at which that amplitude occurs (see an example plot in Figure~\ref{fig:critical_velocity}). These two values can be used to compute a peak velocity $v_{sb}$, which should fall onto the backbone curve described by equation \ref{eq:backbone}. When $v_{s,b}^2$ is plotted against $1- \omega_b^2/\omega_{00}^2$, we expect the relationship to a straight line with a slope proportional to $v_c^2$. There exists a drive regime where the measured amplitudes begin to deviate from this backbone curve due to the neglected term in equation \ref{eq:u_full} (see the discussion in Appendix D). We focus in this paper on the drive regime where the data is well described by the Duffing-like model, which we call the weakly non-linear regime.

\section{Results}

\begin{figure}[t]
    \centering
    \includegraphics[width=\linewidth]{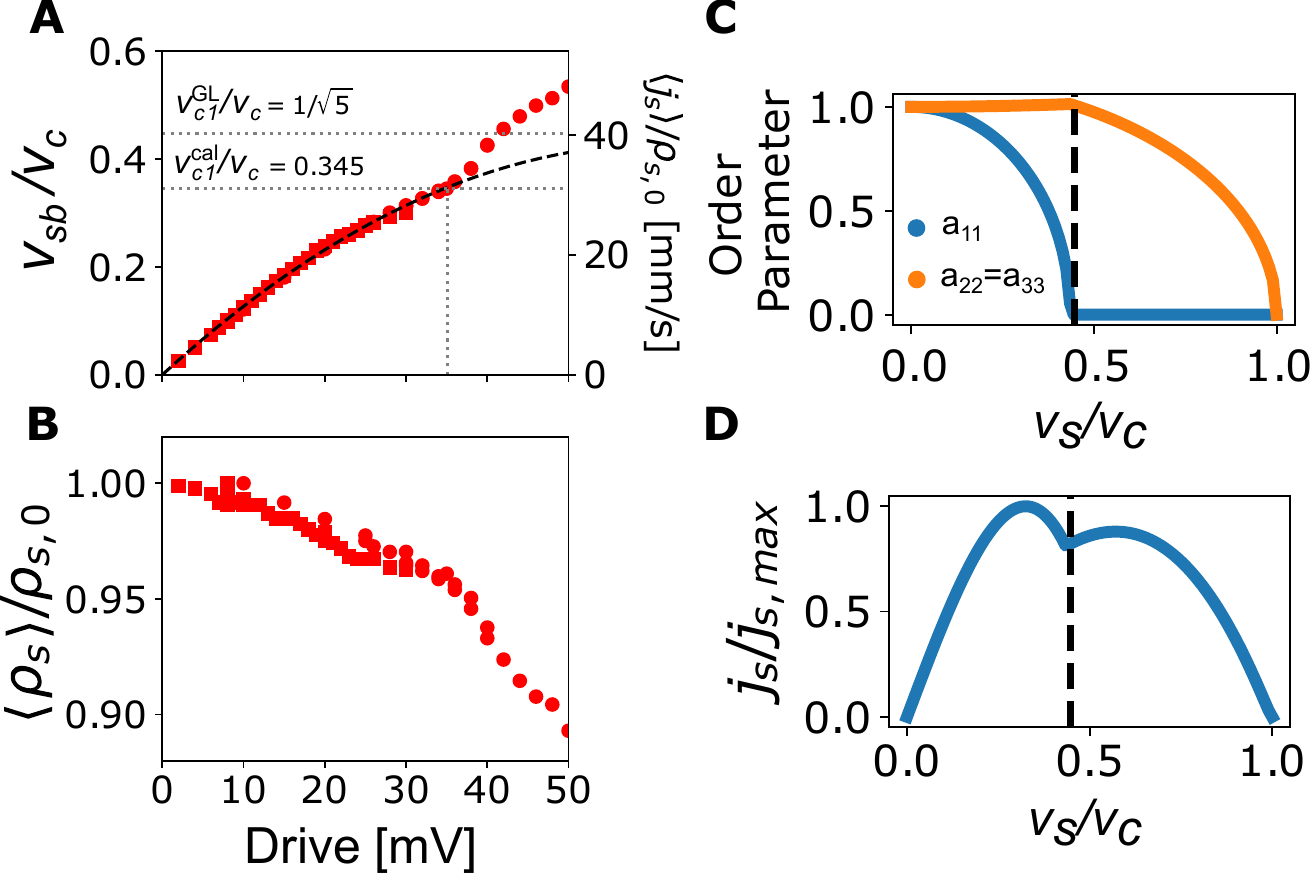}
    \caption{a) Plot of the calibrated superfluid velocity as a function of drive voltage at 25 bar and $T=0.72 T_c$ for the 1800 nm device. Two data sets are superimposed (one plotted as squares, the other circles), which were taken on different dates but are otherwise identical. A discontinuity in the slope can be identified at a critical velocity $v_{c1}^{\textrm{cal}} = 31$ mm/s, which in normalized units natural to the GL theory is $(2m^* \xi/\hbar) v_{c1}^{\textrm{cal}} = 0.345$. b) Plot of the superfluid fraction calibrated from the resonant frequency. A change in slope is observed at the same critical velocity. 
    c) A plot from a  Ginzburg-Landau simulation performed for an 1800 nm parallel plate geometry at 25 bar and $T=0.72 T_c$ of the order parameter components $a_{11}(z=d/2)$ (blue), and $a_{22}(z=d/2) \approx a_{33}(z=d/2)$ (orange) at the center of the slab (here only $a_{22}$ is plotted as the difference is too small to resolve). The component $a_{11}(z=d/2)$ can be seen to go to zero at the critical velocity $v_{c1}^{\textrm{GL}}/v_c = 1/\sqrt{5}$ (marked by the dashed black line), which represents a phase transition from a B-like phase to a planar-like phase. d) Plot of the normalized mass current as a function of normalized velocity calculated using the same GL model.}
    \label{fig:critical_velocity}
\end{figure}

By converting the peak voltage of the Duffing-like resonances into a velocity, we can study features of the drive scaling. Figure~\ref{fig:critical_velocity} 
 shows a representative data set for the 1800 nm measurement at 25 bar and a temperature of $T=0.72 T_c$. Figure~\ref{fig:critical_velocity}a shows the drive scaling of the amplitude measurement. Here, we have chosen to normalize our calibrated velocity by $v_c = \hbar/2m^*\xi$ to facilitate comparison to the GL theory. The quantity $\rho_{s,0}$ is determined by measuring the low drive resonant frequency, then computed using Equation~\ref{eq:resonance_freq}. The frequency at which the maximum occurs is used to compute the velocity-dependent superfluid density (see Equation~\ref{eq:rhos_vs_scaling}) as shown in Figure~\ref{fig:critical_velocity}b. In the drive scaling of both $v_{sb}/v_c$ and $\langle \rho_s \rangle/\rho_{s,0}$, we identify a clear critical velocity, $v_{c1}$, where the slope changes. This critical velocity is well outside the linear regime of the Helmholtz resonator, implying that it cannot be the velocity at which bound state excitations are pumped into the bulk as in References~\cite{castelijns1986landau,bradley2016breaking,zheng2017critical,shook2024surface}. Although the scaling described in Equation~\ref{eq:rhos_vs_scaling} is sufficient to predict the Duffing-like resonance, it does not predict an intermediate critical velocity between the velocity where pair breaking first occurs and when pair breaking is total.
 
 To understand the appearance of this additional critical velocity, we consider a Ginzburg-Landau model based on the work of Fetter and Ullah~\cite{fetter1988superfluid}. The details of the derivation can be found in Appendix A. We are aware of other models of gap suppression due to the superfluid flow \cite{vollhardt1978depairing,vollhardt1980anisotropic,kopnin1992superflow,buchholtz1993superfluid3he}, which may be able to better describe low temperatures and capture the effect of Fermi liquid corrections, but we choose to make use of the GL model because of its conceptual simplicity. The theory expands the thermodynamic free energy in terms of the superfluid order parameter, which for superfluid $^3$He-B flowing between parallel plates has a matrix form
 \begin{equation}
 A =
     \begin{pmatrix}
         A_{11} & 0 & 0 \\
         0 & A_{22} & 0 \\
         0 & 0 & A_{33}
     \end{pmatrix},
 \end{equation}
 which describes the magnitude of the superfluid gap for a particular direction in momentum space. The bulk system is isotropic with $a_{11}=a_{22}=a_{33}$, but the existence of walls suppresses one or more components. For rough surfaces where scattering is diffuse (expected in our experiment), all components go to zero at the surface and recover their bulk values at a few coherence lengths away from the surface \cite{li1988superfluid}. We define the highly confined direction as the $z$-direction, which is associated with $a_{33}$, and take the flow to be in the $x$-direction that is associated with $a_{11}$. For zero flow, the $x$ and $y$ directions (which are both macroscopically large) are indistinguishable, such that $a_{11} = a_{22} \neq a_{33}$, which corresponds to the order parameter of the planar distorted B-phase. For finite flow, the GL theory predicts suppression primarily of the $a_{11}$ component (with a much smaller degree of suppression in $a_{22}$, and $a_{33}$), up until the velocity at which $a_{11} = 0$. Beyond this velocity $a_{22}$ and $a_{33}$ are suppressed (see Figure~\ref{fig:critical_velocity}c). This critical velocity at which $a_{11} = 0$ represents a phase transition to a planar-like phase \cite{saraj2025dimensional}. The dimensionless velocity at which this occurs is
 \begin{equation}
     \left(\frac{v^{\textrm{GL}}_{c1}}{v_c}\right)^2 = \frac{1-2\zeta_{12}}{\gamma} \approx \frac{1}{5},
 \end{equation}
 where $\zeta_{12}$ is a normalized version of GL $\beta$-parameters (defined in Appendix A) which in weak-coupling theory is $\zeta_{12} = 1/5$. 
 
 The parameter $\gamma$ describes the ratio of the gradient parameter in the theory, which for the weak coupling theory is $\gamma = 3$ in bulk $^3$He. There is reason to believe that this parameter may change somewhat in the confined system \cite{saraj2025dimensional}, but the extent is not known, so we consider the bulk value. Strong coupling corrections are known to introduce temperature and pressure dependence to $\zeta_{12}$; however, the effect on the critical velocity is small and likely less significant than other simplifying assumptions. We therefore elect to use the weak-coupling values.
 
\begin{figure}[t]
    \centering
    \includegraphics[width=\linewidth]{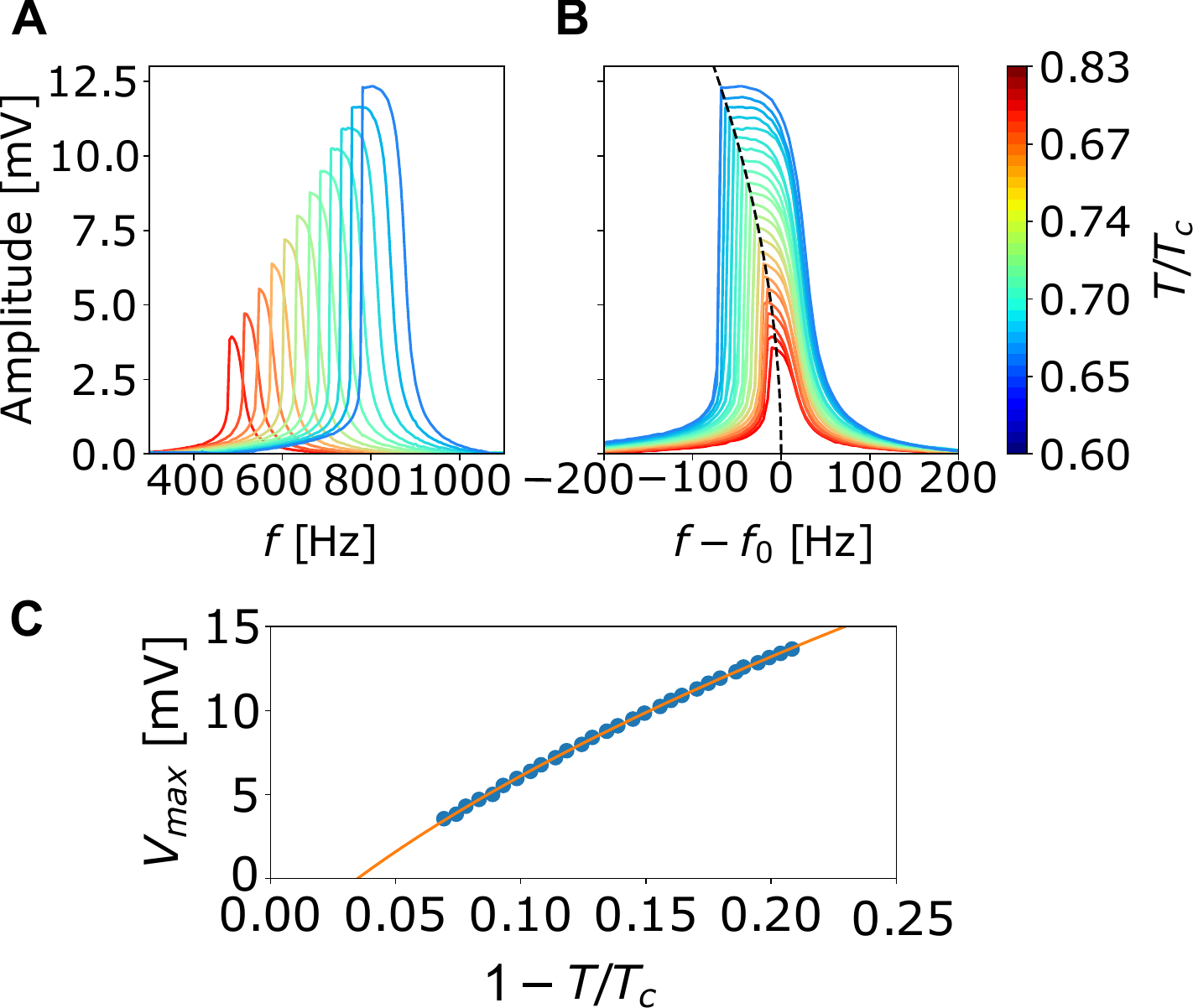}
    \caption{a) The Helmholtz resonance for the 750 nm device is plotted for the reverse sweep direction at a fixed drive voltage (50 mV) for a range of temperatures at 15.7 bar. The resonance shifts up in frequency due to the temperature dependence of the thermodynamic (i.e., velocity-independent) superfluid density $\rho_{s,0}(T)$. The shape of the curve can also be seen to change, becoming more strongly non-linear at lower temperatures. This is due to the temperature dependence of the critical velocity changing the relative scale of the Duffing non-linearity. b) The same resonances are shown, but shifted by subtracting the center frequency $f_0$. The colored boxes mark the point at which the frequency drops discontinuously due to the switching between two branches. For a Duffing oscillator, this point coincides with the maximum of the curve. Here, some distortion of the Duffing-like line shape can be observed for the lower temperatures, due to the neglected term in Equation~\ref{eq:u_full}. Despite this, the frequency discontinuity points can be fit to a quadratic function, shown as a dashed black curve.}
    \label{fig:temperature_scaling}
\end{figure}

In Figure~\ref{fig:critical_velocity}, we measure $v^{\textrm{cal}}_{c1}/v_c = 0.345$, which is a deviation from the model by $(v^{\textrm{GL}}_{c1} - v^{\textrm{cal}}_{c1})/v^{\textrm{GL}}_{c1} = 0.229$. Given the simplicity of the model, this discrepancy does not seem unreasonable. We note that the Ginzburg-Landau model is not expected to be quantitatively accurate for temperatures far from $T_c$, however, the qualitative aspect of $a_{11}$ going to zero at a critical velocity, $q^{\textrm{GL}}_{c1}$, should hold at all temperatures. To investigate the temperature dependence of the non-linear resonance, we performed a temperature sweep measurement with constant drive voltage, shown in Figure~\ref{fig:temperature_scaling}. Here, one can see two effects. There is a shift in the resonance due to the temperature dependence of $\rho_{s,0}(T)$, which is consistent with what we see in the linear regime. This effect can be subtracted out by plotting the resonance as a function of frequency detuning $f-f_0$, where $f_0$ is the linear regime resonant frequency. The second effect is the temperature dependence of the amplitude, which is proportional to $\rho_s$. This is analogous to changing the drive and results in resonances that are more non-linear at lower temperatures. At the lowest temperatures, the shape of the resonance begins to deviate from the parabolic backbone of the Duffing oscillator. This is due to an interaction between the non-linear line shape and the frequency dependence of the bridge measurement described in the Supplementary material.

Figure \ref{fig:temperature_scaling} shows the temperature scaling of the maximum voltage measured for the reverse sweep peak. Here, the temperature is calibrated based on an estimation of $\omega_{00}$ from the forward sweep direction. Because the forward sweep curve resembles a Lorentzian far from resonance, it is possible to identify a central line at $\omega_{00}$ around which the curve is symmetric for frequencies not too close to $\omega_{00}$. This does not exactly coincide with the maximum value of the forward sweep, but differs on the order of the linewidth. Our expectation is that the amplitude curve should be proportional to $v_c \propto \sqrt{1-T/T_c}$ based on the derivation in Appendix C. We find that the data fits a similar curve, but with a small offset in temperature. The discrepancy may be due to a systematic error in our estimation of $\omega_{00}$.




\section{Conclusion}

We have made measurements of the hydrodynamic properties of superfluid $^3$He in channels where one dimension is highly confined. We observe a non-linear regime where the resonance becomes bistable and study the drive dependence. The scaling is found to be well described by a model where a cubic term is introduced to the equation of motion due to the velocity dependence of the superfluid density. In addition to the bistability, we observe a critical velocity within the non-linear regime, which we attribute to the longitudinal component of the order parameter being suppressed to zero. We compare the observed critical velocity to a simple Ginzburg-Landau model and find reasonable agreement. Lastly, we made measurements of the temperature dependence of the critical velocity and found it to be consistent with the model.

Characterization of the velocity dependence of $\rho_s$ in the Helmholtz resonator devices is an important stepping stone in using these devices as probes for subtle effects such as the modification of $\rho_s$ due to surface-bound states described in \cite{wu2013majorana}. Notably, implementing a future Helmholtz resonator device capable of testing the proposal of Wu and Sauls \cite{wu2013majorana} would involve superimposing a DC mass flow on top of the fourth sound resonance, adding further complexity. Typically, it is desirable to operate the devices near the linear regime; however, the ability to account for non-linear corrections allows for the investigation of more subtle effects. Comparison of the hydrodynamic model developed to describe the Helmholtz resonators to a simple Ginzburg-Landau model both yields a novel critical velocity measurement and provides a test of the validity of the hydrodynamic models.

\begin{acknowledgments}
The authors acknowledge that the land on which this work was performed is in Treaty Six Territory, the traditional territories of many First Nations, Métis, and Inuit in Alberta. They acknowledge the support from the Canada Foundation for Innovation; the Natural Sciences and Engineering Research Council, Canada (Grant Nos.~RGPIN-2022-03078,  ALLRP 602120-2024); and Alberta Innovates (Grant No.~242506367).
\end{acknowledgments}

\appendix

\section{Ginzburg-Landau Model}

Here, we consider a superfluid $^3$He system confined between parallel plates with surface normal in the $z$ direction, which is flowing in the $x$ direction. Following Fetter and Ullah \cite{fetter1988superfluid}, we assume that the form of the order parameter is
\begin{equation}
    A(x,z) = \Delta_B e^{iqx} \begin{pmatrix} 
    a_{11}(z) & 0 & 0 \\
    0 & a_{22}(z) & 0 \\
    0 & 0 & a_{33}(z)
    \end{pmatrix},
\end{equation}
where $\Delta_B$ is the bulk B-phase gap, and
\begin{equation}
    q = \frac{2m}{\hbar} v_s,
\end{equation}
is the dimensionless superfluid velocity. We note that Buchholtz \cite{buchholtz1993superfluid3he} suggests including imaginary $a_{13}$ and $a_{31}$ components; however, we find this does not modify the result of the critical velocity calculation, so we chose to use the simplest possible model. Using the formalism described in \cite{rudd2021strong}, we write the normalized bulk free energy as
\begin{equation}
\begin{split}
    f_B = \frac{F_B}{\alpha \Delta_B^2} & = -(a_{11}^2+a_{22}^2+a_{33}^2) \\
    & + \frac{1}{2} \zeta_{12} (a_{11}^2+a_{22}^2+a_{33}^2)^2 \\
    & + \frac{1}{2} \zeta_{345} (a_{11}^4+a_{22}^4+a_{33}^4),
\end{split}
\end{equation}
where $\alpha$ is the second-order Ginzburg parameter, and $\zeta_i$ are the fourth order parameters normalized according to
\begin{equation}
    \zeta_{i} = \frac{\beta_i}{3\beta_{12}+\beta_{345}}.
\end{equation}
The bulk gap $\Delta_B$ is related to these parameters according to
\begin{equation}
    \Delta_B = \sqrt{\frac{\alpha}{2(3\beta_{12}+\beta_{345})}}.
\end{equation}
The normalized gradient free energy is 
\begin{equation}
\begin{split}
    f_G = \frac{F_G}{\alpha\Delta_B^2} & = q^2(\gamma a_{11}^2 + a_{22}^2 + a_{33}^2) \\
    & + \left( a_{11}' \right)^2 + \left( a_{22}' \right)^2 + \gamma \left( a_{33}' \right)^2.
\end{split}
\end{equation}
Here the primes denote derivatives in $z$. All spatial distances are defined in units of the coherence length $\xi = \sqrt{K/\alpha}$, where $K$ is the gradient Ginzburg-Landau parameter. Typically three parameters are defined, $K_{1,2,3}$, however we make the standard simplifying assumption that $K_2=K_3=K$ and $K_1=(\gamma - 1)K$. In weak coupling theory $\gamma=3$ for the bulk system. Strong coupling corrections will increase this parameter slightly to $\gamma \sim 3.1$, however, a more substantial modification due to confinement should suppress this parameter to the range $1 \leq \gamma \leq 3$ \cite{saraj2025dimensional}.

To solve for the spatial dependence of the order parameter components, we solve the associated Euler-Lagrange equation for each component. This gives three equations of motion
\begin{equation}
    a_{11}'' = (\gamma q^2 - 1) a_{11} + \zeta_{12}(a_{11}^2+a_{22}^2+a_{33}^2)a_{11} + \zeta_{345} a_{11}^3,
\end{equation}
\begin{equation}
    a_{22}'' = (q^2 - 1) a_{22} + \zeta_{12}(a_{11}^2+a_{22}^2+a_{33}^2)a_{22} + \zeta_{345} a_{22}^3,
\end{equation}
\begin{equation}
    \gamma a_{33}'' = (q^2 - 1) a_{33} + \zeta_{12}(a_{11}^2+a_{22}^2+a_{33}^2)a_{33} + \zeta_{345} a_{33}^3.
\end{equation}
For this calculation, we will assume diffuse scattering conditions such that all components of the order parameter must go to zero at the walls. The first critical velocity that appears in this set of equations is the value of $q$ for which $a_{11}(z)=0$ everywhere. To identify this value, we solve the system of equations numerically using the SciPy module solve$\_$bvp to iterate an initial guess solution. A natural guess for the order parameter components is an elliptical function of the form
\begin{equation}
    a_{ii} = a_i \textrm{sn}\left(\left.\frac{\pi z}{d} \right| k_i \right),
\end{equation}
where sn is a Jacobi elliptic function, $k_i$ is the elliptic modulus, $a_i$ is the amplitude, and $d=D/\xi$ is the confined dimension in units of coherence length. When the order parameter is a scalar value, a Jacobi elliptic function of this form is the exact solution of the GL equations with the boundary conditions considered here. From prior experience computing order parameters for the planar distorted B-phase, we know that $a_{33}$ typically has a spatial distribution very close to sn$(z)$, and the other components are similar, except close to the walls. This makes the function sn$(z)$ an appropriate initial guess.

From the numerical calculations, we find that the first critical velocity closely coincides with the function
\begin{equation}
    q_{c1}^2 = \left( \frac{v_{c1}^{GL}}{v_c} \right)^2 = \frac{1-2\zeta_{12}}{\gamma}.
\end{equation}
For weak coupling values $\zeta_{12} = 1/5$, and $\gamma=3$ such that $q_{c1} = 1/\sqrt{5}$.

\section{Non-linear Oscillator Equations}

The equations specifying the amplitude of the Helmholtz oscillator may be written as 
\begin{equation}
    \tilde{\omega}^4 + b \tilde{\omega}^2 + c = 0,
\end{equation}
where
\begin{equation}
    b = \tilde{\nu}^2 - \frac{\tilde{G}_0^2}{x_0^2} - 2\left(1 - \frac{3}{4}x_0^2 \right), \quad c = \left(1 - \frac{3}{4}x_0^2 \right)^2,
\end{equation}
and $\tilde{\omega} = \omega/\omega_0$, $\tilde{\nu} = \nu/\omega_0$, $\tilde{G}_0 = G_0/\omega_0$, are the normalized frequencies. The solutions $\tilde{\omega}(x_0)$ are given by
\begin{equation}
    \tilde{\omega}_{\pm}^2(x_0) = \frac{-b\pm\sqrt{b^2-4c}}{2}.
\end{equation}
This solution is real when $b^2-4c > 0$. The maximum amplitude is attained at the point where the two branches meet at $b^2-4c=0$. Multiplying each side of this relation by $x_0^4$, expanding, then factoring the common term, yields the condition for the maximum amplitude $x_m$
\begin{equation}
    \left(\tilde{\nu}^2x_m^2 - \tilde{G}_0^2 \right) \left[ x_m^4 + \frac{(\tilde{\nu}^2-4)}{3} x_m^2 - \frac{1}{3} \tilde{G}_0^2 \right] = 0
\end{equation}
which has the solutions
\begin{equation}
    x_{m,1,2} = \pm \frac{G_0}{\nu},    
\end{equation}
\begin{equation}
    x_{m,3,4} = \pm \sqrt{\frac{(\tilde{\nu}^2-4)}{3} + \frac{1}{2}\sqrt{\frac{(\tilde{\nu}^2-4)^2}{9} + \frac{4}{3} \tilde{G}_0^2}},
    \label{eq:max_velocity_scaling}
\end{equation}
\begin{equation}
    x_{m,5,6} = \pm \sqrt{\frac{(\tilde{\nu}^2-4)}{3} - \frac{1}{2}\sqrt{\frac{(\tilde{\nu}^2-4)^2}{9} + \frac{4}{3} \tilde{G}_0^2}}.
\end{equation}

\section{Temperature Dependence of Critical Temperature}

The critical velocity at which $a_{11}$ goes to zero predicted by the Ginzburg-Landau model is 
\begin{equation}
    v_{c1}(T) \approx \frac{v_c(T)}{\sqrt{5}} = \frac{1}{\sqrt{5}} \frac{\hbar}{2m^*\xi(T)}, 
\end{equation}
where $v_c$ is the critical velocity at which all components of the order parameter go to zero. The temperature dependence in the model comes from the Ginzburg-Landau coherence length
\begin{equation}
    \xi(T) = \frac{1}{\sqrt{1-T/T_c}} \sqrt{\frac{7\zeta(3)}{20}} \frac{\hbar v_F}{2\pi k_BT_c},
\end{equation}
in which case
\begin{equation}
    v_{c}(T) \approx \frac{\pi k_B T_c}{m^*v_F} \sqrt{\frac{20}{7\zeta(3)}} \sqrt{1 - \frac{T}{T_c}}.
\end{equation}
The scaling of the maximum resonance amplitude for the Duffing-like regime is shown in Appendix B in Equation~\ref{eq:max_velocity_scaling}. When the line width is small compared to the center frequency, such that $\tilde{\nu}^2 = (\nu/\omega_{00})^2 \ll 1$ and $\tilde{G_0} = G_0/\omega_{00} \gg 1$, there is a simple square root scaling
\begin{equation}
    x_{m,3} = \frac{v_{sm}}{v_{c}} 
    \approx \sqrt{\frac{\tilde{G}_0}{2\sqrt{3}}},
\end{equation}
\begin{equation}
    \Rightarrow v_{sm} = \frac{\pi k_B T_c}{m^*v_F} \sqrt{\frac{10}{7\sqrt{3}\zeta(3)}} \sqrt{\tilde{G}_0(1-T/T_c)}.
\end{equation}

\section{Highly Non-linear Regime}

Here, we consider the case where the force term in equation 23 cannot be neglected. Using the drive $F_E(t) = F_0 \sin(\omega t)$, and ansatz $v_s = v_0 \cos(\omega t + \theta)$, the signal is 
\begin{equation}
    u(t) = \frac{\omega F_0}{k_p D} \cos(\omega t) + \frac{\omega^2 \rho_0 A \ell}{k_p D} v_0(\omega) \cos(\omega t + \theta). 
\end{equation}
This signal is then demodulated by the lock-in amplifier to produce the quadratures
\begin{equation}
    X(\omega) = \frac{1}{T} \int^T_0 u(t) \cos(\omega t) dt = \frac{\omega F_0}{k_p D} + \omega^2 \frac{\rho A \ell}{k_p D} v_0 (\omega),
\end{equation}
\begin{equation}
    Y(\omega) = \frac{1}{T} \int_0^T u(t) \sin(\omega t) dt = - \omega^2 \frac{\rho A \ell}{k_p D} v_0(\omega).
\end{equation}
The amplitude quadrature then, is given by
\begin{equation}
\begin{split}
    R^2(\omega) & = X^2(\omega) + Y^2(\omega) \\ & = \left( \frac{\omega F_0}{k_p D} \right)^2 \left[1 + 2\omega \rho A \ell \frac{v_0}{F_0} + 2 (\omega \rho A \ell)^2 \frac{v_0^2}{F_0^2} \right].
\end{split}
\end{equation}
This results in a distortion in the shape of the non-linear resonance, which appears as a rounding of the top of the resonance.

\bibliography{bib.bib}

\end{document}